\documentclass[12pt,a4paper,twoside]{article}
\usepackage{epsfig}
\usepackage{amssymb,latexsym,amsfonts}
\usepackage{times}
\usepackage{bm}

\newcommand{\be}{\begin{equation}}
\newcommand{\ee}{\end{equation}}
\newcommand{\bd}{\begin{displaymath}}
\newcommand{\ed}{\end{displaymath}}
\newcommand{\ba}{\begin{array}}
\newcommand{\ea}{\end{array}}
\newcommand{\bt}{\begin{tabular}}
\newcommand{\et}{\end{tabular}}

\newcommand{\bea}{\begin{eqnarray}}
\newcommand{\eea}{\end{eqnarray}}

\newcommand{\Z}{\mathbb{Z}}

\newcommand{\R}{\mathbb{R}}

\newcommand{\mod}{\textrm{mod}}
\newcommand{\id}{\mathbf{1}}
\newcommand{\im}{\mathrm{i}}

\begin{document}
\begin{flushright}
INLO-PUB-06/00\\
\end{flushright}

\vspace{.5cm}

\begin{center}
{\Large Orientifolds and twisted boundary conditions}

\vspace{.5cm}

{\bf Arjan Keurentjes}\footnote{email address: \sl arjan@lorentz.leidenuniv.nl} \\
{\it Instituut-Lorentz for theoretical physics, Universiteit Leiden \\ P.O. Box 9506, NL-2300 RA Leiden, 
The Netherlands}\end{center}

\vspace{.5cm}

\begin{abstract}
It is argued that the T-dual of a cross-cap is a combination of an $O^+$ and an $O^-$ orientifold plane. Various theories with cross-caps and D-branes are interpreted as gauge-theories on tori obeying twisted boundary conditions. Their duals live on orientifolds where the various orientifold planes are of different types. We derive how to read off the holonomies from the positions of D-branes in the orientifold background. As an application we reconstruct some results from a paper by Borel, Friedman and Morgan for gauge theories with classical groups, compactified on a 2-- or 3--torus with twisted boundary conditions.
\end{abstract}

\section{Introduction}

The advent of D-branes and orientifolds in string theory gave new tools to study gauge theories. The vacua of gauge theories with classical groups, compactified on tori with commuting holonomies can be straightforwardly described by configurations of D-branes and orientifold planes (see \cite{Polchinski} and references quoted there). For theories with unitary or symplectic groups, holonomies are specified in the fundamental representation, for theories with orthogonal groups the holonomies are in the vector representation. All these representations still have a non-trivial centre. 

When fields are invariant under the centre, typical for conventional open string perturbation theory, one can allow for holonomies that commute up to an element of the centre, as was first considered for gauge theory by 't Hooft \cite{Hooft} in terms of so-called twisted boundary conditions. But it was only relatively recent that this extra freedom was first considered in the context of string theory. In \cite{Witten97} Witten studied the case of $SO(4N)$ on 2-- and 3--tori, which can be described by a configuration of D-branes on an orientifold.

In the appendix of the same paper, Witten used a construction involving D-branes on an orientifold to show that for orthogonal groups the moduli space of flat connections on the 3--torus with periodic boundary conditions has an extra component not considered before, and that this seems to solve an old problem concerning the computation of the Witten index \cite{Witten82}. Motivated by this, various authors \cite{Borel} \cite{Kac} \cite{Keurentjes98} \cite{Keurentjes99a} \cite{Keurentjes99b} have subsequently shown the necessary existence of extra vacuum components for exceptional groups, so as to solve the Witten index problem for these groups, for which no D-brane construction is available. The authors of \cite{Borel} also included the case of general boundary conditions, likewise demonstrating a richer structure than considered earlier in the computation of the Witten index with twisted boundary conditions \cite{Witten82}. It has remained a challenge to translate all these results into configurations of D-branes and orientifold planes, which allowed Witten to make his discovery for the orthogonal groups \cite{Witten97}. Here we close the circle by addressing this translation for all classical groups, with arbitrary boundary conditions.

The fact that in standard open string perturbation theory, all representations are conjugate to the adjoint (and therefore invariant under the centre) is believed to be false outside perturbation theory \cite{Sen98a} \cite{Sen98b} \cite{Witten98}. It is argued that the full gauge group is actually $Spin(32)/\Z_2$, and that also spinorial representations occur. But even configurations that would be consistent for $Spin(32)/\Z_2$-gauge theory, can be shown to be inconsistent for the string theory by more subtle arguments \cite{deBoer}. However, in this paper we wish to elucidate the underlying gauge group structure, which we stress is interesting in its own right. It is also an essential step towards a cleaner, and more complete derivation of the string consistency conditions governing orientifolds. We should note however, that there will be important modifications to our results on allowed gauge groups and representations in orientifold compactifications in $D < 10$ upon imposition of the string consistency conditions.

The main part of this paper is devoted to a discussion of compactification with orthogonal and symplectic groups on 2-- and 3--tori, with twisted boundary conditions (as previously analysed in \cite{Borel} \cite{Schweigert}), but we will also discuss $U(n)$ theory with various boundary conditions. After one T-duality, these theories correspond to configurations of branes and orientifold fixed planes on the M\"obius strip, the Klein bottle, and tori that are not rectangular. We derive how to T-dualise the M\"obius strip and Klein bottle in the direction orthogonal to the first T-duality. This leads to orientifolds with fixed planes of different type, much like in \cite{Witten97}. With these methods every possible flat connection for symplectic gauge groups or orthogonal gauge groups allows a translation in terms of a configuration of D-branes on an orientifold. Holonomies and possible enhanced symmetry groups are easily read off from the configuration.

We will open with a discussion of a $U(n)$-theory on a 2--torus with special boundary conditions: Along one of the directions of the 2--torus there is a holonomy that is not an element of $U(n)$.

\section{The T-dual of a cross-cap}\label{crosscap}

We start by considering $U(n)$ theory on a circle. $U(n)$ can be embedded in an open string theory by attaching Chan-Paton charges on the ends of oriented strings. Compactifying this string theory on a circle and applying a T-duality transformation, we obtain a configuration of $n$ D-branes that are transverse to a dual circle, each intersecting the dual circle in one point. The location of the D-branes is controlled by the holonomy $\Omega_1$ along the circle in the original theory. We are interested in configurations with discrete symmetries. The discrete symmetries of the circle are the shift symmetries, shifting the circle by an angle $2 \pi q$ with $q$ a rational number, and the order $2$ reflection symmetry. Only for specific choices of the original holonomy will the D-brane configuration respect one or some of these symmetries. In this section, we are interested in the reflection symmetry.

The reflection on the circle has two fixed points, which will be taken to be at $X=0$ and $X=\pi R$ ($X$ being the coordinate along the circle, and $2 \pi R$ its circumference). This can always be arranged: in the original theory we had a holonomy in $U(n)$, which is locally equivalent to $U(1) \times SU(n)$. The holonomy for the $U(1)$-factor can be chosen arbitrarily, since it does not couple to anything. In the dual theory this corresponds to an overall translation, which we use to set the coordinates of the fixed points to the above values.

\begin{figure}[h] \label{Klein}
\begin{center}
\includegraphics[height=4cm]{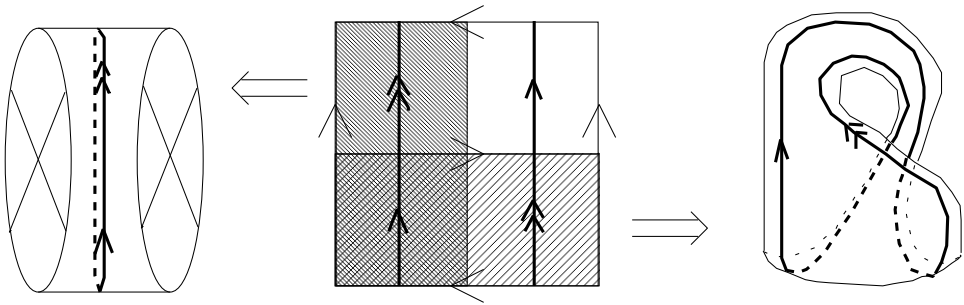}
\caption{The Klein-bottle: a double cover of the Klein-bottle, arrows indicating the direction of identifications (middle); an attempt to draw the standard representation of the Klein-bottle, obtained by taking the lower half of the double cover as fundamental domain (right); the cylinder with two cross-caps, obtained by taking the left half of the fundamental domain (left). We also drawn an example of a brane in all three pictures (depicted twice on the double cover) as it is positioned after the first T-duality}
\end{center}
\end{figure}

Now compactify in addition on another circle of radius $R'$ with holonomy $\Omega_2$ along this circle. The standard formalism assumes holonomies that can be diagonalised within the group. In case we have the above $\Z_2$ symmetry, we may consider a holonomy that includes the $\Z_2$ reflection. Gluing the circle to a reflected circle upon going around the second cycle, one does not obtain a 2--torus, but a Klein bottle. Instead of a non-trivial line bundle over a circle, we will represent the Klein bottle here as a cylinder of length $\pi R$, circumference $4 \pi R'$, bounded by two cross-caps at the end, the cross-cap being an identification over half the period of the circle. 

The D-branes are wrapped around the cylinder, parallel to the cross-caps. There are two possibilities, controlled by the holonomy $\Omega_1$ in the original theory. D-branes in the bulk (away from the cross-caps) represent branes that were reflected into their image. In this representation, D-brane and image are represented as one brane (which is in this sense a brane pair). In the original theory there can also be D-branes at the fixed point(s) of the $\Z_2$-reflection. In this representation of the Klein-bottle, they are located at the cross-cap. Under a smooth deformation of the original holonomy, only even numbers of D-branes can move away from the cross-cap. Hence for $U(n)$ with $n$ odd there is at least one brane fixed under the $\Z_2$ reflection and therefore stuck to a cross-cap. For $n$ even there are two possibilities: the number of branes at each cross-cap is either even or odd. In the latter situation there is at least one brane at each cross-cap.

The above is reminiscent of the situation for orientifold planes. For orientifolds, a brane and an image brane on the double cover are mapped to a brane-pair in the orientifold. There is also the possibility of single branes being stuck at an orientifold plane (in the case of $O^-$ planes. By $O^-$ we denote the orientifold plane that gives orthogonal gauge symmetry, and $O^+$ is an orientifold plane that gives symplectic gauge symmetry).

The above configuration can be interpreted in terms of the original gauge theory. $U(n)$ is locally $U(1) \times SU(n)$, and the $U(1)$ background is fixed. For $n > 2$, $SU(n)$ possesses an outer automorphism, which, in a suitable representation, corresponds to complex conjugation. We will be working in the fundamental representation, and denote complex conjugation as $C$ with action
\be
C \quad : \quad  U \ \rightarrow  \ U^{*} \qquad U \ \in \ SU(n)
\ee
One can extend this action to $U(n)$ as $C$ also has a simple action on $U(1)$, and now one may also extend to $n \leq 2$. One normally considers holonomies taking values in the gauge group, which corresponds to combining a translation in space with the action of an inner automorphism (i.e. a conjugation) on the group. One may also consider a holonomy that corresponds to an outer automorphism, and this is precisely what we are doing in the above. The outer automorphism $C$ can be combined with an inner one, say conjugation with a group element $A$. To avoid ambiguities we require that $AC=CA$, which is true if $A$ is real, that is $A \ \in \ O(N)$.  The holonomy $\Omega_2$ combines the action of $C$ with conjugation with $A$, and we denote it as $\Omega_2 = AC$, with $A$ in the fundamental representation of $U(n)$, and $C$ the operator that implements complex conjugation. The holonomy $\Omega_1$ is an ``ordinary'' holonomy, and we write $\Omega_1=B$, with $B$ an element of $U(n)$ in the fundamental representation. $\Omega_1$ should commute with $\Omega_2$, which is solved by taking $B$ commuting with $A$ and $B \ \in \ O(N)$ . Continuous variation of the $U(1)$-background is incompatible with complex conjugation; in the D-brane picture this corresponds to the fact that a global translation on the D-branes is incompatible with the reflection for generic cases. 

By conjugation with $O(n)$ matrices, we may transform $A$ and $B$ to a block diagonal form with $2 \times 2$ blocks of the form 
\be \label{SO(2)}
\left( \ba{rr} \cos \phi & - \sin \phi \\ \sin \phi & \cos \phi \ea \right)
\ee
on the diagonal, and some $1$'s and $-1$'s as remaining diagonal entries. In the following, we will take $A$ and $B$ to be of this standard form.

We wish to T-dualise the cylinder with the two cross-caps in the direction of the circle. Ignoring the cross-caps one would roughly expect this to lead to a dual theory on a cylinder. The inclusion of the cross-caps can be analysed by examining the symmetries of the original theory. 

$A$ and $B$ are elements of the vector representation of $O(n)$, and in particular their eigenvalues occur in pairs: If $\exp \im \phi$ is an eigenvalue, then so is $\exp -\im \phi$. The ordering is unimportant as there are symmetries that allow the exchange of $\exp \im \phi$ and $\exp -\im \phi$, for every $\phi$ separately. If $A$ and $B$ where holonomies for an $O(n)$-theory in an orientifold description this symmetry would be simply the orientifold projection itself. This suggests that also for this $U(n)$-theory the dual should be some orientifold.

The radius of the dual theory is expected to be $1/(2R')$, half the ``normal'' radius. The coordinates of the D-branes in this theory reflect the eigenvalues of the holonomies in the original theory. Naively mapping these onto the dual circle suggests a circle of radius $1/R'$, which seems to lead to a contradiction. The resolution to this paradox lies in the presence of the operator $C$. If $\Omega_i$ are the holonomies for a certain theory, then the holonomies $\Omega_i' = g\Omega_ig^{-1}$ with $g$ some element of $U(n)$ represent the same theory. Consider the set of diagonal matrices with entries $\pm 1$, $\pm \im$ on the diagonal that commute with the $A$ and $B$. Taking $g$ to be a specific element from this set has the effect  
\bd
g \quad : \quad  (B, AC)  \ \rightarrow  \ (gBg^{-1}, gACg^{-1}) = (B,A g^2 C).
\ed
This leaves $\Omega_1$ and $\Omega_2$ in standard form, but with $A$ replaced by $A g^2$. Hence in this construction, $A$ and $Ag^2$ have to be identified. $g^2$ is an element of $O(n)$ that commutes with $A$. By a suitable choice of $g^2$, any eigenvalue $\exp \im \phi$ of $A$ can be mapped to $-\exp \im \phi= \exp \im (\phi + \pi)$. Therefore the periods of the circle and the eigenvalues of the holonomies match, and the dual theory is indeed an orientifold. Now we examine the orientifold planes.

To find maximal symmetry groups we set $B$ to either $\id$ or $-\id$. If we set $A = \id$, then the surviving symmetry group is the subgroup of $U(n)$ that is invariant under $C$, which is $O(n)$. For another maximal symmetry group, assume $n$ to be even for a moment and take $A$ to be of block diagonal form with $2 \times 2$ blocks
\be \label{Sp}
\left( \ba{rr} 0 & 1 \\ -1 & 0 \ea  \right),
\ee
on the diagonal, and call this matrix $J$. The unbroken symmetry group is then the subgroup of $U(n)$ of matrices $U$ that commute with  $JC$. $C$ transforms $U \rightarrow U^{*}$, but as $U$ is unitary, $U^{*} = (U^{-1})^T$, where $T$ is for transposed. We may then rewrite the invariance condition to 
\be \label{symp}
U^{T} J U = J
\ee
which, together with the unitarity condition defines the symplectic group. It is obvious how to generalise to arbitrary  holonomies, and odd $n$: a holonomy with $k$ blocks $\textrm{diag}(1,1)$ and $k'$ blocks (\ref{Sp}) gives rise to $O(k) \times Sp(k')$-symmetry, completed with some $U(m)$-factors, whenever $m$ eigenvalues not equal to $\pm 1$, $\pm \im$ coincide.

The above $U(n)$ theory on a Klein-bottle is thus T-dual to an orientifold $T^2/\Z_2$, where two of the four orientifold planes are of $O^-$-type and two are of $O^+$-type. The holonomy $B=\pm \id$ distinguishes two parallel configurations of one $O^+$ and one $O^-$-plane, whereas in the theory on the Klein bottle it distinguished the two parallel cross-caps. As a rule of thumb one may therefore state that the dual of the cross-cap is a configuration of one $O^+$ and one $O^-$-plane. This fits with the usual charge assignments: opposite charges for the $O^+$ and $O^-$ plane versus no charge for the cross-cap. The original theory may have had isolated D-branes at the cross-caps. In the dual theory the isolated branes should be located at the $O^-$-planes, since the $O^+$ planes cannot support isolated branes. Examining the holonomies that will lead to such a situation indeed shows this to be the case.

These ideas are independent of whether D-branes are static in the background, or used as ``probes''. The above orientifold background is identical to a IIB-orientifold encountered in \cite{Witten97}, but consistency requires absence of D-branes. This suggests to regard this model as a ``$U(0)$-theory'' with a holonomy that includes complex conjugation. Its duality to IIA on a Klein-bottle is obvious from the above. Considering various limits one may also reach other theories discussed in \cite{Witten97} and \cite{Dabholkar}. 

We interpreted the Klein-bottle theory as created by combining a translation with an outer automorphism (complex conjugation). Outer automorphisms can always be divided even if not combined with a translation. Dividing a $U(n)$ group by its outer automorphism will give a symplectic or orthogonal theory, where the ambiguity comes from the fact that an outer automorphism may be combined with an inner automorphism to give another outer automorphism. In our case one may consider, instead of $C$, an operator $AC$ with $A$ an element of $U(n)$. One should require $A$ to commute with $C$ and therefore $A \ \in O(n)$. Consistency also requires that $AC$ acting on the group squares to the identity. The group action on itself is always in the adjoint representation, and hence we have the possibilities $A^*A = (A^{-1})^T A = \pm \id$, so $A$ is either symmetric or antisymmetric. One may now copy a standard textbook derivation \cite{Polchinski} to show that this leads to either symplectic or orthogonal groups. The reasoning is parallel to that for orientifolds, so one may interpret the introduction of an orientifold plane as quotienting the gauge group by an outer automorphism. With this point made, which is not stressed in the literature, we may also say that in the above a translation is combined with an orientifold action, as the theory in the last chapter of \cite{Witten97} was originally motivated. 

\section{Twisted boundary conditions on the 2--torus}
\subsection{Twist in unitary groups}

The $U(n)$-gauge group allows a second form of twist. The circle also has discrete shift symmetries by angles $2 \pi q R$, with $q$ a rational number, which can be chosen on the interval $[0,1)$. Choose a configuration of the $n$ D-branes that respects one or some of these shifts. In that case the number $q$ is a multiple of $1/n$. Now compactify on a second circle with a holonomy that includes the shift over $2 \pi q R$. This results in a theory on the 2--torus, not with $n$ D-branes, but with $k = \textrm{gcd}(qn, n)$ branes, wrapped $n/k$ times around the torus. This theory is naturally interpreted as a $U(n)$-theory with twisted boundary conditions \cite{Hooft}. We will not have much new to say on this theory, but mention it for completeness, and to point out some effects that are encountered in other theories as well.

Let $X_1$ and $X_2$ be the coordinates transverse resp. parallel to the branes. Then this 2--torus is $\R^2$ with coordinates $(X_1, X_2)$, quotiented by a lattice generated by the vectors
\be
e_1 =  2 \pi(q R_1,  R_2)  \qquad e_2 = 2 \pi ( R_1, 0) 
\ee
Now transform to an $SL(2, \Z)$-equivalent form. Let $n' = n/k$. Then $n'$ and $qn'$ are integer, and $\textrm{gcd}(qn', n') = 1$. Hence the equation 
\bd
n'a + qn' b = 1, \qquad a,b \ \in \Z
\ed 
has a solution, which can be found using Euclid's algorithm. The solution is not unique as $a \rightarrow a + m qn'$, $b \rightarrow b - m n'$ with integer $m$ gives another solution. Use this arbitrariness to select a $b$ such that $ 0 \leq b < n'$. Then change the fundamental domain of the torus by using the $SL(2, \Z)$ transformation
\be
\left( \ba{c} x' \\ y' \ea \right) = \left( \ba{cc} n' & b \\ -qn' & a \ea \right) \left( \ba{c} x \\ y \ea \right),
\ee
where $(x,y) \ \in \ \Z$ are coordinates for the lattice vectors $xe_1 + ye_2$. Under this transformation the basis vectors transform as 
\be
 2 \pi ( qR_1, R_2) \rightarrow 2 \pi (0,  n'R_2) \qquad
 2 \pi ( R_1, 0 ) \rightarrow  2\pi ( R_1/n', bR_2 )  
\ee
On this fundamental domain only $k$ D-branes (which are in a sense configurations of $n'$-tuples of branes) are visible. This is analogous to the two different representations of the Klein bottle in the previous section. The set-up is the one considered in \cite{Douglas}, which is argued to lead to a Yang-Mills theory on a non-commutative torus in a suitable limit. 

We may T-dualise our original theory back to an open string theory (with Neumann boundary conditions) on a torus, using the standard methods \cite{Giveon}. The resulting theory has a non-zero $B$-field (with $B=q$ in appropriate units) in the background, as our original theory does not live on a square torus. 

Combining the discrete shift over $2 \pi q R$, with the $Z_2$ reflection does not lead to anything new. The resulting transformation is of the form
\bd
X \rightarrow -X + 2 \pi q R
\ed
which is just a $\Z_2$-reflection, but with other fixed points. This is related to the $U(n)$ theory of the previous section by a trivial translation. 

\subsection{Twist in symplectic groups on the 2--torus}

Symplectic groups can be realised in string theories by combining the Chan-Paton construction with the gauging of world sheet parity \cite{Polchinski}. This gives a theory of unoriented strings. To complete the description of the theory one has to prescribe how world sheet parity acts on the Chan-Paton matrices. If the reflection of the world sheet is combined with the action of an anti-symmetric matrix on the Chan-Paton indices, the resulting theory will have symplectic gauge symmetry.

Compactifying this theory on a circle of radius $R_1$ and T-dualising leads to an oriented string theory, living on an interval $I= S^1/ \Z_2$ of size $(2R_1)^{-1}$, bounded by two $O^+$-planes. For an $Sp(k)$-theory there will be $k$ D-brane pairs distributed along the interval. The two $O^+$-planes do not allow any freely acting shift. We will instead assume that the D-branes are distributed in a configuration that is invariant under the reflection that exchanges the two $O^+$-planes. For odd $k$ one brane-pair is fixed in the middle of the interval. Now compactify on another circle of radius $R_2$ with a holonomy that implements the $\Z_2$-reflection. The resulting compactification manifold is a M\"obius strip with an $O^+$-plane as edge. For $k$ even half of the D-branes are exchanged with the other half on going around the circle. For $k$ odd half of $(k-1)/2$ pairs are exchanged with another $(k-1)/2$ and one brane-pair is fixed by the $\Z_2$ reflection. Another representation of the M\"obius strip, is a cylinder of diameter $2R_2$ and length $(4R_1)^{-1}$. One end of the cylinder ends in a cross-cap, the other end is formed by a single $O^{+}$-plane. On the cylinder there are $k/2$ D-branes pairs for $k$ even, and $(k-1)/2$ for $k$ odd in which case there is a brane-pair stuck at the cross-cap. 

\begin{figure}[h] \label{Mobius}
\begin{center}
\includegraphics[height=4cm]{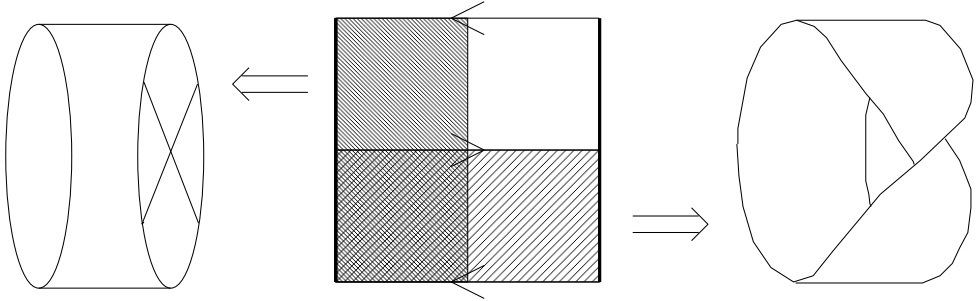}
\caption{The M\"obius strip: a double cover of the M\"obius strip, arrows indicating the direction of identifications, fat lines the edges (middle); the standard representation of the M\"obius strip, obtained by taking the lower half of the double cover as fundamental domain (right); the cylinder with one cross-cap, obtained by taking the left half as fundamental domain (left)}
\end{center}
\end{figure}

This theory is a $U(2k)$-theory as described in section \ref{crosscap} with an extra orientifold plane inserted. The mirror symmetry of the orientifold plane turns the Klein bottle into a M\"obius strip. Take the circle that is dual to the circle of radius $R_1$ and choose coordinates as follows: we will take the orientifold planes at $X=0$ and $X= \pi /R_1$, and the fixed points of the $\Z_2$-reflection at $X = \pi /2 R_1$ and $X = 3 \pi/ 2 R_1$. The description from the $U(n)$ theory has to be slightly modified, as the fixed points of the $\Z_2$ are no longer located at $X=0$ and $X= \pi R$, as before. The action of the $\Z_2$-reflection can be interpreted in the original theory as accomplished by the operator $(-\id) C$, which is complex conjugation combined with multiplying by $(-\id)$. In symplectic theories one projects onto states invariant under $JC$, with $J$ the matrix composed of $2 \times 2$-blocks of the form (\ref{Sp}), and the invariance condition is (\ref{symp}). In the orientifold projected theory, the operator $(-\id) C$ is identified with $(-\id) \textrm{Ad}{J}$, which has as action ``conjugate with $J$ and multiply with $-\id$''. Multiplying by $-\id$ is not an outer automorphism of $Sp(k)$ (in fact, the symplectic groups do not posses any outer automorphism at all), and it can be realised by conjugation, as we will show later.

With the appropriate symmetries realised, we can pass from the $U(n)$-theory to the symplectic theory as follows. We argued that the $U(n)$-theory had as its holonomies $(\Omega_1, \Omega_2) = (B, AC)$. Replace the operator $C$ by $(-\id)C$, and then perform the orientifold projection. The resulting holonomies are then $(\Omega_1, \Omega_2) = (B, A(-\id)\textrm{Ad}J)$. $\Omega_1$ and $\Omega_2$ do not commute, but anticommute. Their eigenvalues can be read of from $B$ and $A$, but we have to find a way to implement the action of $-\id$.

Anticommutativity of the holonomies is allowed in symplectic theories, provided all representations of $Sp(k)$ have trivial centre (this is the case for all representations one encounters in $Sp(k)$ string perturbation theory. These are the adjoint, which is the symmetric two-tensor; a $k(2k-1)-1$ dimensional representation which is the antisymmetric tensor with an extra singlet removed; and the singlet). This theory may also be analysed by the methods of \cite{Schweigert}. Here we will reproduce the results of such an analysis by a different method.  

The T-dual theory to the M\"obius strip is an orientifold $T^2/\Z_2$, with the size of the $T^2$ being $(2R_1)^{-1} \times (2R_2)^{-1}$ (one fourth of the usual size, compare with \cite{Witten97}). At the four fixed points we find orientifold fixed planes. The original $O^+$-plane splits into two $O^+$-planes intersecting the torus at a point. The cross-cap will dualise into one $O^+$-plane and one $O^-$-plane, so we have a total of 3 $O^+$-planes and 1 $O^-$-plane. On the dual we have $k/2$ D-brane pairs at arbitrary positions if $k$ is even. If $k$ is odd, there are $(k-1)/2$ D-branes whose positions can be chosen freely. The remaining brane pair was stuck at the cross-cap, so in the dual picture there is an isolated brane at an orientifold plane, which should be the $O^-$.

The corresponding holonomies can be read of as follows. A brane pair in the bulk has two coordinates, and each corresponds to four eigenvalues $\lambda_i, -\lambda_i, \lambda_i^{-1}, -\lambda_i^{-1}$ with $\lambda_i = \exp (2 \pi \im X_i / R_i)$, $X_i$ and $R_i$ being the coordinate and the radius of the corresponding dimension (the $O^-$ plane is located at $(X_1/R_1, X_2/R_2) = (1/4,1/4)$, the remaining $O^+$ at $(0,0)$, $(1/4,0)$, $(0,1/4)$). Corresponding to these eigenvalues we have $2 \times 2$ blocks on the diagonals of the holonomies of the form
\be
\left( \ba{rr} \lambda_1 & 0 \\ 0 & -\lambda_1 \ea \right) \qquad \left( \ba{rr}  0 & -\lambda_2 \\ -\lambda_2 & 0 \ea \right)
\ee 
where the left block appears in one of the holonomies and the other, resulting from multiplying a diagonal block with a block of the form (\ref{Sp}) in the other holonomy. There is a second set of blocks with $(\lambda_1, \lambda_2)$ replaced by $(\lambda_1^{-1}, \lambda_2^{-1})$. For a single brane located at the $O^-$ plane we get blocks with $(\lambda_1, \lambda_2) = (\im, \im)$. One easily verifies that this prescription leads to anticommuting elements in the fundamental representation of the symplectic group. 

On the orientifold $T^2/\Z_2$ one should introduce a $B$-field which is half-integer valued. For orthogonal groups this is well known, and it is usually deduced from a path-integral argument \cite{Sen97}. It may also be deduced from duality. The M\"obius strip we used may be described as the torus $T^2$ quotiented by the lattice generated by $2 \pi(0, 2R_2)$ and $2 \pi (R_1/2, R_2)$, quotiented by an orientifold action that takes $(X_1, X_2) \rightarrow (-X_1, X_2)$. Omitting the orientifold for a moment, we see that the torus is skew, implying a half-integer value for the $B$-field in its dual \cite{Giveon}. The same reasoning applies to a M\"obius strip, where the edge is formed by an $O^-$ instead of $O^+$ -plane. This corresponds to an orthogonal theory without vector structure, as described in \cite{Witten97}, and reproduced by our analysis later. 

The resulting orientifolds describe the moduli space of compactifications of $Sp(k)$ theories with twisted boundary conditions. As a check consider the cases $k=1$ and $k=2$, since as $Sp(1)/\Z_2 = SU(2)/\Z_2 = SO(3)$, and $Sp(2)/\Z_2 = SO(5)$ these results should be reproduced by other orientifolds. $Sp(1)$ with twist corresponds to an orientifold with 3 $O^+$-planes, and a single D-brane stuck to the $O^-$ plane. The resulting configuration allows no continuous gauge freedom, in accordance with the standard description of $SU(2)$ with twist. The single D-brane at the $O^-$ fixed plane gives $O(1) = \Z_2$ residual symmetry; this should be interpreted as the symmetry of the centre of $SU(2)$ which is the only symmetry of $SU(2)$ that survives the twist.

For $k=2$ the dual description consists of a single D-brane-pair on the orientifold with 3 $O^+$ and 1 $O^-$-planes. The rank of the unbroken group is 1, and generically it is $U(1)$. At the $O^-$-plane this is enhanced to $O(2)$, while at any of the three $O^+$-planes it is enhanced to $Sp(1) = SU(2)$. We will see in the next section that this nicely agrees with the orientifold description of the $O(5)$ orientifold corresponding to the $\Z_2$-twisted case. 

For higher $k$ the analysis is similar. For $k$ even the generic unbroken group is $U(1)^{k/2}$, which can be enhanced to $U(k/2)$ at a generic position at the orientifold, to $Sp(k/2)$ at one of the three $O^+$ planes or $O(k)$ at the $O^-$ point. For $k$ odd this analysis can be copied while replacing $k$ by $k-1$, with the exception that at the $O^-$ $O(k)$ symmetry is possible because of the brane already present there.

\subsection{Twist in orthogonal groups on the 2--torus}

Twist in the orthogonal groups gives a more involved situation and we can distinguish several possibilities. Every orthogonal group has a two-fold cover, so the resulting $Spin$-group has at least a $\Z_2$ centre. Compactification on a two torus with twist in this $\Z_2$ will lead to absence of ``spin-structure'': fields in the spin representation are not allowed since the holonomies will not commute in this representation. 

For $SO(N)$-theories with $N$-odd this is all one can do apart from compactification with periodic boundary conditions. For $N$ even, $SO(N)$ already has a non-trivial centre and the above mentioned $\Z_2$ is just a subgroup of the whole centre. For $N$ divisible by $4$, the centre of $Spin(N)$ is $\Z_2 \times \Z_2$. $\Z_2 \times \Z_2$ allows three $\Z_2$ subgroups (basically each of the $\Z_2$ factors, and a diagonal embedding). Of these two are related by the outer automorphism of the $Spin(N)$-groups with $N$ even. Hence there are two options for twisting by a $\Z_2$: The already above mentioned $\Z_2$ leading to compactification without spin structure, and a second one, named ``compactification without vector structure''. The latter is named so because in this compactification the vector representation is not an  allowed one.

For $Spin(N)$ with $N$ even but not divisible by $4$, the centre is $\Z_4$. The previously mentioned $\Z_2$ is generated by the order $2$ element in $\Z_4$. It is also possible to twist by an element of $\Z_4$ generating the whole centre. We will call this compactification without vector structure, since in this case the vector representation is not an allowed one either. Note however that this twist forbids any representation with a non-trivial centre, so the spin representation should be absent as well. The only representations allowed in this case are conjugate to the adjoint. The results from this section may also be derived with the methods of \cite{Schweigert}. We will follow a different route.

\subsubsection{No spin structure}

Absence of spin-structure does not forbid the vector representation, so one can use an ordinary orientifold $T^2/\Z_2$ with 4 $O^-$-planes. The topological non-triviality has to be treated with the technique of Stiefel-Whitney classes, following the appendix of \cite{Witten97}. Absence of spin-structure implies that the second Stiefel-Whitney class is non-vanishing. We will keep on demanding that the first Stiefel-Whitney class vanishes (which implies $SO(N)$-symmetry, not only $O(N)$), and hence the total Stiefel-Whitney class should be $w= 1 + \omega_2$, with $\omega_2$ the 2-form on the 2-torus. For $SO(N)$ with $N$ odd, this is accomplished by placing 3 branes at the three non-trivial orientifold fixed points, placing $(N-3)/2$ pairs at arbitrary points. For $SO(N)$ with $N$ even one distributes 4 branes over all orientifold fixed points and has $(N/2 -2)$ pairs at arbitrary locations. These are the only solutions.

It is instructive to compare the cases of $SO(3)$ and $SO(5)$ without spin structure to the analysis for $Sp(1)$ and $Sp(2)$ with twist, as $Spin(3) = Sp(1)$ and $Spin(5) = Sp(2)$. For $SO(3)$, 3 isolated D-branes are located at three orientifold planes, and there is no continuous gauge freedom at all, just like in the $Sp(1)$ case. There are discrete symmetries $O(1)^3 = \Z_2^3$, but these just correspond to the $8$ diagonal $O(3)$-matrices. Of these, 4 are not elements of $SO(3)$, and of the remaining 4, 3 lift to elements that anticommute with the holonomies in $Spin(3) = Sp(1)$. Only the identity remains, which lifts to the two centre elements of $Sp(1)$, which is how the $\Z_2$ discrete symmetry there is recovered.

For $SO(5)$, we have 3 orientifold planes occupied by one brane each, and a pair of branes at an arbitrary point. Generically the unbroken symmetry is $U(1)$, which can be enhanced to $O(3)$ at any of the three points where an orientifold plane with brane is sitting. At the remaining orientifold plane $U(1)$ is enhanced to $O(2)$. Stressing again that $SO(3)$ is the double cover of $Sp(1)$, we see that this is exactly the same as the $Sp(2)$ orientifold with twist.

Further easy examples are $SO(4)$ and $SO(6)$ without spin structure. For $SO(4)$ there is no residual gauge symmetry. The $\Z_2$ associated with vector structure acts as twist in both $SU(2)$-factors of $Spin(4)$, eliminating all gauge freedom. $SO(6)$ without spin structure is equivalent to $Spin(6) = SU(4)$ with $\Z_2$-twist. From the above description this gives a rank 1 subgroup, which can be enhanced to $O(3)$ at 4 orientifold-planes. This coincides with the $SU(4)$-description, where $SU(2)$ is a maximal symmetry group. 

\subsubsection{No vector structure}

The case of absence of vector structure was already analysed by Witten \cite{Witten97} for $O(4N)$. Here we present an analysis from a different point of view, which also nicely extends to the case of $O(4N+2)$.

Compactifying a string theory with orthogonal gauge symmetry $SO(2k)$ on a circle, and T-dualising along this circle, gives a theory on the interval $I= S^1/ \Z_2$. The interval is bounded by two $O^-$-planes, and on the interval we have $k$ pairs of D-branes. We will assume that the $O^-$ planes do not contain any isolated D-branes, since if both of them would be occupied we do not have $SO(2k)$ but $O(2k)$-symmetry, and if only one of them would be occupied this would indicate $O(n)$-symmetry with $n$ odd (actually, for $n$ odd the following construction is impossible, which is a reflection of the fact that $SO(n)$ with $n$ odd allows only one kind of twist).

Again the only possible discrete symmetry is $\Z_2$ reflection symmetry, and we will henceforth assume that this is realised. Compactifying on an extra circle, with a holonomy implementing this reflection leads again to a theory on the M\"obius strip, this time with $O^-$planes on the boundary. We again go to the representation in which the M\"obius strip is a cylinder, bounded on one end by an $O^-$ plane, and on the other end by the cross-cap. Notice that for $k$ odd, there is a pair of D-branes fixed by the reflection and, on the cylinder it has to be located at the cross-cap. Half the number of the remaining D-branes are visible in this representation. Since we are restricting to $SO(n)$-configurations, we can repeat the whole discussion presented for symplectic groups, with the difference that the orientifold planes we insert here will not give symplectic but orthogonal symmetry. From the geometric picture one may again deduce anticommutativity of the holonomies $\Omega_1$ and $\Omega_2$. 

We may now T-dualise as before, and obtain an orientifold with two $O^-$-planes from the original $O^-$-plane, and an $O^+$ and $O^-$-plane from the cross-cap. This explains the relation between the IIA-theory on a M\"obius strip, that is discussed in \cite{Park}, and the IIB orientifold in \cite{Witten97}. Both feature in a network of theories describing duals of the CHL-string \cite{Chaudhuri95a}. The strong coupling limit of IIA theory on the M\"obius strip gives M theory on a M\"obius strip, which has another weak coupling description as a heterotic $E_8 \times E_8$ string, with a holonomy interchanging the two $E_8$-factors \cite{Chaudhuri95b}. Via heterotic duality, this corresponds to a compactification of the $Spin(32)/\mathbb{Z}_2$-string without vector structure as described in \cite{Lerche} \cite{Witten97}. In the latter paper, the strong coupling description of the heterotic $Spin(32)/\mathbb{Z}_2$-string without vector structure is derived to be the IIB-orientifold with a single $O^+$ and 3 $O^-$-planes. The $\mathbb{Z}_2$ projection causing the reduction of rank of the gauge group is geometrical in the IIA and M theory descriptions (compare with similar constructions in \cite{Schwarz} \cite{Chaudhuri95c} \cite{Chaudhuri95d}). The T-duality derived here closes the circle, and relates the mechanism for rank reduction in the IIB-theory to the more transparant one of the IIA-theory.

We may represent the parts of the holonomies corresponding to branes in the bulk in the same way as in the symplectic case. These can be conjugated to matrices in the real vector representation of $O(n)$. Readers who prefer the real representation of $O(n)$ should substitute for each brane in the bulk the following $4 \times 4$-blocks
\be
\left( \ba{cccc} 
\cos \phi_1 & 0 & -\sin \phi_1 & 0 \\
0 & -\cos \phi_1 & 0 & \sin \phi_1 \\
\sin \phi_1 & 0 & \cos \phi_1 & 0 \\
0 &- \sin \phi_1 & 0 & -\cos \phi_1 \ea \right) \qquad
\left( \ba{cccc} 
0 & -\cos \phi_2 & 0 & \sin \phi_2 \\
-\cos \phi_2 & 0 & \sin \phi_2 & 0 \\
0 & -\sin \phi_2 & 0 & -\cos \phi_2  \\
- \sin \phi_2 & 0 & -\cos \phi_2 & 0 \ea \right) 
\ee
as can be derived straightforwardly. The parameters $\phi_i =  X_i/2 \pi R_i$ follow from the coordinates of the D-branes on the torus.

For $k$ odd, there was an odd number of D-brane pairs at the cross-cap, and hence the $O^-$-plane coming from the cross-cap has to contain an odd number of branes. The other two orientifold points also contain isolated branes. This is possible because the even number of branes that should be on the edge of the M\"obius strip (to ensure $SO(2k)$-symmetry), may translate into odd numbers of branes at each of the corresponding $O^-$-planes in the dual theory.   

Consider a single brane stuck to the $O^-$-plane that came from dualising the cross-cap. As can be seen from the geometric picture, this corresponds to a block $\textrm{diag}(\im, -\im)$ in the holonomy $\Omega_1$, or equivalently, a block
\bd 
\left( \ba{rr} 0 & -1 \\ 1 & 0 \ea \right)
\ed  
in the real representation of $O(n)$. Demanding anticommutativity with a $2 \times 2$ block in the holonomy $\Omega_2$ leads to the unique solution $\textrm{diag}(1,-1)$ (in the real representation, up to conjugation with an element of $SO(2)$). One may also consider a single brane stuck to one of the other $O^-$ planes. This defines a block $\textrm{diag}(1,-1)$ in the holonomy $\Omega_1$. Demanding anticommutativity with a second $2 \times 2$ block leads to two inequivalent possibilities, being
\bd 
\left( \ba{rr} 0 & -1 \\ 1 & 0 \ea \right) \qquad
\left( \ba{rr} 0 & 1 \\ 1 & 0 \ea \right)
\ed 
These two possibilities correspond to the two $O^-$-planes that came from dualising the original $O^-$-plane. The point is now that occupying 1 or 2 of the $O^-$ planes by an odd number of branes corresponds to holonomies in $O(n)$, but occupying all three at once with an odd number of single branes does give holonomies in $SO(n)$. This also gives the interpretation for the orientifold with 3 $O^-$ planes and one $O^+$-plane where not all of the $O^-$ planes are occupied; these represent $O(k)$ configurations that cannot be represented in $SO(k)$. We see that if we demand $SO(n)$-symmetry, and occupy one $O^-$-plane with an odd number of branes, we have to occupy all $O^-$ planes by an odd number of branes.

On the resulting dual orientifold we have $k/2$ pairs of D-branes if $k$ is even, or $(k-3)/2$ if $k$ is odd. For $k$ even we obtain back the description of \cite{Witten97}, with the possibility of $O(k)$ symmetry at 3 planes, and $Sp(k/2)$-symmetry at 1 plane. For $k$ odd we have the possibility of $O(k-2)$ at three planes, and $Sp((k-3)/2)$ at one plane.  

It is again instructive to look at a few examples. $k=1$ is impossible in $SO(2)$. $k=2$ corresponds to $SO(4)$-theory without vector structure. Since $Spin(4)$ is $SU(2) \times SU(2)$, this corresponds to twist in one of the $SU(2)$ factors, and arbitrary holonomies in the other $SU(2)$-factor. In the orientifold description, we have the possibility of enhanced symmetries $O(2)$ and $Sp(1)$. $Sp(1)=SU(2)$ obviously corresponds to the unbroken second factor. The $O(2)$'s correspond to situations in which the holonomies in the second factor are $(1,\im \sigma^3)$, $(\im \sigma^3,1)$, $(\im \sigma^3,\im \sigma^3)$. The extra parity transformation is due to the fact that $\im \sigma^3$ anticommutes with the elements $\im \sigma^{1,2}$, which lifts to the double cover as commutation symmetry.

$k=3$ gives $SO(6)$ without vector structure. This corresponds to $Spin(6) = SU(4)$ with $\Z_4$-twist. The absence of remaining gauge freedom is completely in agreement with the $SU(4)$ description.

$k=4$ gives $SO(8)$ without vector structure, which due to triality should be equivalent to $SO(8)$ without spin structure. We certainly do find $Sp(2) = Spin(5)$ symmetry in both descriptions. $O(4)$ is less visible in the above description of $SO(8)$ without spin structure, but this is due to the fact that the parity in $O(4)$ is actually not a real symmetry (compare to the $O(2)$ symmetry found for $k=2$), and $Spin(4) = SU(2) \times SU(2)$ can also be obtained outside orientifold fixed planes. The asymmetry in the two description is then due to the fact that they represent different projections from the same moduli space.

\section{Compactifications on a 3--torus}
\subsection{Commuting triples for orthogonal and symplectic groups} \label{3commute}

For compactifications on a 3--torus with periodic boundary conditions, the vacua are classified by its three holonomies, which should be 3 commuting elements in the gauge group.

Finding such a triple for a symplectic theory amounts to placing a number of D-brane pairs on an orientifold $T^3/\Z_2$ where the fixed points are all $O^+$-planes. A similar thing can be done for orthogonal theories, with the difference that on the fixed points of the $\Z_2$ orientifold action one introduces $O^-$-planes, which can also support single D-branes. Not all such configurations correspond to flat connections with periodic boundary conditions. Alternative boundary conditions can be described by this orientifold, as long as the holonomies commute in the vector representation of $O(N)$. For a periodic connection one should however demand that the holonomies commute in $Spin(N)$, which is a more severe restriction. To solve which configurations of $O(N)$ holonomies lift tusefulo $Spin(N)$ holonomies, one may calculate the Stiefel-Whitney class for the configurations \cite{Witten97}. Only if the Stiefel-Whitney class is trivial the configuration can be lifted to $Spin(N)$.

Solving for these requirements Witten found new vacua for orthogonal gauge theory with periodic boundary conditions on a 3--torus \cite{Witten97} whose existence was explained from the group theory point of view in \cite{Borel} \cite {Kac} \cite{Keurentjes98} \cite{Keurentjes99a} \cite{Keurentjes99b}. A configuration with 7 D-branes on the 7 fixed points excluding the origin of an orientifold, and $N$ pairs at arbitrary positions parametrises a periodic connection for $SO(2N+7)$-theory. Similarly, a configuration with 8 D-branes distributed over all 8 fixed points, and $N$ pairs at arbitrary positions parametrises a periodic connection of $SO(2N+8)$-theory. Besides these there are always the flat periodic connections that are smoothly connected to the configuration where all three holonomies are equal to the identity. For $SO(N)$ with $N$ even, this gives a configuration of only D-brane pairs on the orientifold. For $N$ odd there is a single isolated D-brane at the origin of the orientifold. Forgetting about the D-brane pairs for a moment, we see that for each theory there are two solutions, one in which a number $k$ of orientifold planes is occupied by single branes ($k$ being $0$ or $1$) and one where $(8-k)$ orientifold planes are occupied by single branes. This is a useful observation for later.   

\subsection{$c$-triples for symplectic groups}

For symplectic groups on the 3--torus one may also choose non-periodic boundary conditions. This can be done in various ways, but by using $SL(3,\Z)$ transformations on the torus, all possibilities are isomorphic to one standard form. We can choose the standard form to have twist between the holonomies in the $1$ and $2$ direction, and the third holonomy commuting with the former two. Following \cite{Borel}, we call a triple of such holonomies a $c$-triple, where $c$ denotes that the three holonomies only commute up to a (non-trivial) centre element of the gauge group.

From our analysis for twist in symplectic gauge theories on the 2--torus, one easily deduces that the corresponding orientifold description has 6 $O^+$-planes and 2 $O^-$planes. The two planes with 3 $O^+$ and 1 $O^-$, are distinguished by the eigenvalue $\pm 1$ in the third holonomy. Eigenvalues for the third holonomy can be read of in the usual way, with the remark that their multiplicities should be doubled. A configuration for the 2--torus may therefore be imported in either of these planes, corresponding to choosing the third holonomy in $Sp(k)$ to be $\pm \id$, which are the two elements of the centre of $Sp(k)$. One quickly deduces that there are always 2 disconnected possibilities for placing the D-branes in this orientifold background.

First suppose $Sp(k)$ symmetry with $k$ even. For the description with twist on a 3--torus, this should give $k/2$ pairs of D-branes in the above orientifold background. There are two possibilities to distribute the D-branes. First, one can have $k/2$ pairs at arbitrary locations on the orientifold. But one can also split one pair, put one D-brane on one $O^-$-plane and the other on the other $O^-$-plane, and have the remaining $(k/2-1)$-pairs at arbitrary locations. Both possibilities are legitimate, since $Svector structurep(k)$ is simply connected. The conclusion is thus, that $Sp(k)$-theory on a 3--torus with twisted boundary conditions has a moduli space of 2 components, one with a rank $k/2$ unbroken gauge group, and one with a rank $k/2-1$ unbroken gauge group. We can now perform a Witten index count for this theory, as also performed in \cite{Borel}: The two components will contribute $k/2+1$ and $k/2$ to the index giving the total value $k+1$ in agreement with both the periodic boundary conditions case, and the infinite volume case \cite{Witten82}.

For $k$ is odd the procedure should also be clear. One can place $(k-1)/2$ pairs of D-branes at arbitrary points in the orientifold background. The single D-brane that is left can go on either of the two $O^-$planes. These are inequivalent possibilities, and hence also in this case the moduli space consists of 2 components. Each of these components contributes $(k+1)/2$ to a Witten index calculation, giving also the correct result $k+1$ \cite{Borel}.

Again we check the $k=1$ and $k=2$ cases. According to the above, $Sp(1)$-theory with twist on a 3--torus gives a moduli space of 2 components. On each component the gauge group is completely broken. This is as it should be as $Sp(1) = SU(2)$ which, when compactified on a 3--torus with twist has a moduli space that looks like this. We again have the remaining $O(1) = \Z_2$ symmetry corresponding to the centre of $SU(2)$, which commutes with everything. 

Perhaps more interesting is the $Sp(2)$-case, where we have one component for which the gauge group is completely broken, and another where a rank 1 group survives. The rank one gauge group is generically $U(1)$, but can be enhanced to $O(2)$ at two planes or $Sp(1)$ at six other planes. This coincides with the description we will find for $SO(5) = Sp(2)/\Z_2$, without spin structure.  

\subsection{$c$-triples for orthogonal groups}

For orthogonal groups on a 3--torus there are more possibilities for the boundary conditions. Like in the case for the 2--torus, we can have absence of either spin- or . By $SL(3,\Z)$-transformations on the torus, we can again arrange that the holonomies for the 1 and 2-direction are the ones that do not commute (in the Spin-cover of the group), while the third holonomy does commute with the other two.

In the case of $SO(4n)$ there is however a new possibility. For $Spin(4n)$ the centre of the gauge group is not cyclic but a product of cyclic groups, being $\Z_2 \times \Z_2$. Call the generator of the first $\Z_2$ $z_s$ (with $s$ for spin), and the generator of the second $\Z_2$ $z_c$ ($c$ being the standard notation for the second spin-representation). Also define $z_v = z_s z_c$. This notation is motivated by the fact that identifying $z_v \sim \id$ gives the vector representation.

We can now also impose the following twist conditions on the holonomies:
\be \label{O4n}
\Omega_1 \Omega_2 = z_s \Omega_2 \Omega_1 \qquad \Omega_2 \Omega_3 = z_c \Omega_3 \Omega_2 \qquad \Omega_3 \Omega_1 = z_v \Omega_1 \Omega_3 
\ee
This can be thought of as a standard form. $SL(3, \Z)$-transformations result in an isomorphic moduli-space. We will call this case ``spin nor vector''-structure, and treat it separately.

\subsubsection{No spin structure}

This is the easiest case, provided we use some previously obtained knowledge. From our description of orthogonal theories on a 2--torus without spin structure, a particular case for the 3--torus can be obtained as follows.

For $SO(k)$ with $k$ odd, place 3 single D-branes at three $O^-$-planes within one plane within the orientifold $T^3/\Z_2$ leaving the fixed plane at the origin empty, and place the others in pairs at arbitrary points at the orientifold. For $k$ even one should place 4 single D-branes at 4 orientifold fixed planes within one plane of $T^3/\Z_2$.

For $k \geq 4$ there is always a second possibility. Remember from section \ref{3commute} and \cite{Witten97} that a configuration of 8 D-branes distributed at all orientifold planes has a trivial Stiefel-Whitney class. We may ``add'' this orientifold configuration to another as follows. Take a specific configuration of D-branes at the orientifold. This has a certain Stiefel-Whitney class, which can be thought of as providing a topological classification for the configuration. Now adding 8 more D-branes at the orientifold fixed points will not affect the Stiefel-Whitney class. This is so because the Stiefel-Whitney class of the 8 D-branes is trivial, and the Stiefel-Whitney class of the "new" configuration may be simply obtained by multiplying the class of the "old" configuration with that of the added configuration (it is important to realise that Stiefel-Whitney classes are $\Z_2$ valued, and that $-1 = 1 \ \mod \ 2$, so there is no ordering ambiguity). One may also add or delete any pair of D-branes without affecting the class, also because of its $\Z_2$ nature.

We thus obtain the following possibilities: For $SO(k)$ with $k$ odd, we had 3 single D-branes at three $O^-$-planes. Adding the 8 D-branes and reducing modulo 2, we obtain a configuration of 5 D-branes with the same topological classification as the previous one. The 5 D-branes are precisely at the orientifold planes that were not occupied previously, and in a sense one could speak of a $\Z_2$-complement. One can add pairs of D-branes to again obtain an $SO(k)$ configuration.

For $k$ even one had 4 single D-branes at 4 $O^-$-planes. Taking the $\Z_2$-complement, we get an inequivalent configuration with 4 single D-branes at the other 4 $O^-$-planes, with the same topological classification. Of course, afterwards we must add pairs of D-branes to acquire $SO(k)$.

We will now discuss several cases.
$k=3$ corresponds to $SO(3)=SU(2)/Z_2$ with twist on the 3--torus. The $SU(2)$ description has two components. The $SO(3)$-description has also two components, but these cannot be distinguished by their holonomies. In a particular representation, the $SU(2)$ holonomies read
\be
\Omega_1 = \im \sigma_3; \qquad \Omega_2 = \im \sigma_1; \qquad \Omega_3 = \pm 1,
\ee
but $\pm 1$ in $SU(2)$ are both projected to the same element of $SO(3)$ being the identity.

$k=4$ gives $SO(4)$ which gives two orientifolds, but some thought will reveal that also in this case there are twice as many components in moduli space. Using that $Spin(4) = SU(2) \times SU(2)$, the no-spin-structure condition amounts to twisting both $SU(2)$ factors simultaneously. For the third holonomy one has then 4 possibilities, being any combination of plus or minus the identity in each $SU(2)$-factor. These 4 possibilities project to only two sets of holonomies in $SO(4)$, and hence two orientifold descriptions.

$k=5$ gives us $SO(5)$ which is interesting because we should be able to reproduce the $Sp(2)$-results here. $SO(5)$ without spin-structure gives two orientifolds. On one we have 5 fixed D-branes and hence no residual gauge symmetry. On the other we have 3 fixed D-branes and a pair wandering freely. Possible enhanced gauge symmetries are $O(3)$ at three points, and $O(2)$ at five points. However, all but one of the ``parity'' symmetries (corresponding to elements with $\det = -1$) in these $O(n)$ groups are ``fake'' in the sense that they correspond to elements that anticommute in $Spin(5)$. The remaining $O(2)$ corresponds to the $O(2)$'s we encountered in the $Sp(2)$ case, and the $SO(3)$'s map to the $Sp(1)$-unbroken subgroups in $Sp(2)$. That the multiplicities of these enhanced symmetry groups are only half of those encountered in the $Sp(2)$ description reflects the fact that $SO(5)$ is a double cover of $Sp(2)$, which also translates to the fact that the moduli space of $Sp(2)$-triples is a double cover of the space of $SO(5)$-triples. The moduli space for the gauge theory is the moduli space of $Sp(2)$-triples, as every set of $SO(5)$ holonomies has two inequivalent realisations in terms of gauge fields. Note however that here the number of components in moduli space agree; the two $SO(5)$-components are double covers of two $Sp(2)$ components, not of 4 $Sp(2)$ components

$k=6$ gives $SO(6)$ whose spin cover is $SU(4)$. Here there are two equal dimension components in moduli space, both with a rank $1$ gauge group which can be enhanced to $SO(3)$ at 4 points. 

It is easy to perform the Witten index count for $k \geq 5$ \cite{Borel}. In these cases we have always two components of the moduli space and two corresponding orientifold representations. For $k$ even, both components contribute $k/2-1$, for a total of $k-2$. For $k$ odd, one component contributes $(k-1)/2$ whereas the second contributes $(k-3)/2$ for a total of $k-2$. Of course these answers are as they should be.

\subsubsection{No vector structure}

From the analysis for the 2--torus we deduce that $O(2k)$ without vector structure on a 3--torus corresponds to an orientifold background with 6 $O^-$-planes and 2 $O^+$-planes. Again eigenvalues for the third holonomy can be read off in the usual way, except that their multiplicities should be doubled. One obvious solution to the boundary conditions is to import the solution for the 2--torus here. 

For $k$ even we have seen that a particular solution is given by placing all D-brane pairs at arbitrary points. For the second solution we take as before the $\Z_2$-complement. We have not defined how the operation of "$\Z_2$-complement" acts on the $O^+$-planes but this is not hard to guess. Since $O^+$ planes cannot support isolated D-branes, they should remain empty. Hence the second solution has all $O^-$ planes occupied by one D-brane, and $k-3$ pairs at arbitrary points. A way to see this is as follows. The smallest group for which the configuration with six isolated branes exists is $SO(12)$. One can take the $SO(6)$ holonomies $\Omega_1$ and $\Omega_2$ that gave ``no vector structure'' on the 2--torus (these are unique up to gauge transformations), to construct the $SO(12)$-holonomies $\Omega_1 \oplus \Omega_1$, $\Omega_2 \oplus \Omega_2$ and $\id \oplus -\id$, where $\id$ stands for the identity in $SO(6)$. That these $SO(12)$-matrices satisfy the required boundary conditions is obvious, and liftings to $Spin(12)$ can be constructed from the liftings of the $SO(6)$-holonomies to $Spin(6) = SU(4)$. That $SO(12)$ is the smallest group allowing these extra solutions can also be deduced with the techniques from \cite{Borel}. 

For $k$ is odd we have 3 $O^-$-planes within one plane occupied by D-branes. A priori one has two possible planes, and actually both give distinct solutions. Note that also in these cases the solutions are each others $\Z_2$-complement. For $k$ odd all components of the moduli space are isomorphic, as also follows form the analysis of \cite{Borel}.

$SO(4)$ without vector structure on a 3-torus gives only one solution, since there are simply not enough D-branes to realise the second one. This gives a rank 1 unbroken gauge group which can be enhanced to $Sp(1)$. A naive calculation of the Witten index would give half of the right answer, but as before, the moduli space consists of 2 components that cannot be distinguished by their holonomies in $SO(4)$. We therefore have to multiply the naive value for the index by two, again obtaining the right answer.

$SO(6)$ without vector structure gives two solutions, but from $SU(4)$-analysis one expects four. Again this is due to the fact that inequivalent solutions exist that cannot be distinguished by their holonomies in $SO(6)$. Notice that the gauge group is completely broken, which is as it should be.

A Witten index calculation is straightforward for these theories. For $SO(4N+2)$ there are always 4 components \cite{Borel} (projected to two orientifolds) that are all isomorphic. Each orientifold has $2N+1$ branes on it, of which $3$ are stuck. The rest should organise in $N-1$ pairs, giving a rank $N-1$ gauge group and contribution $N$ to the Witten index. With 4 components one obtains the total value $4N$, which is indeed the dual Coxeter number for these theories. 

For $SO(4N)$ one has 2 components (1 orientifold), where no branes are stuck and $N$ pairs move freely, giving a contribution of $2N+2$ to the index. For $N \geq 3$ this is not sufficient, but for these cases there exist 2 more components (1 orientifold) having 6 stuck branes, and $N-3$ pairs at arbitrary points. The total adds up to $2(N+1) + 2(N-2) = 4N-2$, the right answer. 

\subsubsection{Spin nor vector structure}

For $SO(4k)$ there is the possibility of holonomies satisfying equation (\ref{O4n}). In some sense this should encompass both the case of no spin- as well as no vector-structure. The orientifold background is as in the case without vector structure, $T^3/\Z_2$ with 2 $O^+$-planes and 6 $O^-$-planes. On top of this $2k$ branes should be distributed.

A clue on the D-brane configuration can be found from T-dualising in the direction of the line connecting the two $O^+$ planes. This gives a theory on the product of a circle and an orientifold $T^2/\Z_2$ with 3 $O^-$'s  and 1 $O^+$ plane. This orientifold corresponded to an orthogonal theory without vector structure on the 2--torus. In our case, the gauge group is $SO(4k)$, and we know that if there are branes at an orientifold plane, their number should be even. Translating back to the orientifold $T^3/\Z_2$, the pair of orientifold planes corresponding to one $O^-$-plane in $T^2/\Z_2$ will be occupied either both by an even number of branes, or both by an odd number of branes. This leaves $8$ possibilities, 2 of which can be quickly discarded as they correspond to a $SO(4k)$-theory without vector, but with spin structure. The remaining possibilities have thus either $2$ or $4$ branes stuck at $O^-$ planes, and hence $2k-2$, resp. $2k-4$ D-branes in the bulk

T-dualising in another direction, along a line connecting an $O^+$ and an $O^-$-plane will lead to an orientifold of the form $((T^2/\Z_2) \times S^1)/\Z_2$. One can represent this by an orientifold $T^2/\Z_2$ with 4 $O^-$ planes, quotiented by a $\Z_2$-reflection in one of the points halfway on the line between 2 $O^-$-planes (the multiple possibilities are related by $SL(2,\Z)$-transformations). This reflection has a second fixed point. Over the orientifold $(T^2/\Z_2)/\Z_2$ one erects a circle everywhere, except at the two fixed points of the $\Z_2$-reflection where it is replaced by a cross-cap. On the orientifold $T^2/\Z_2$ we should have absence of spin structure, meaning that all $O^-$-planes are occupied by an odd number of branes (remember that $4k$ is even). This translates to occupancy of both $O^-$-planes in the quotient $(T^2/\Z_2)/\Z_2$. We now have two possibilities; either there are an odd number of branes at both cross-caps, or there are an even number. For the orientifold $T^3/\Z_2$, these two possibilities translate into the situations with two $O^-$-planes occupied (cross-caps occupied by even number of branes), and four $O^-$-planes occupied (cross-caps occupied by odd number of branes). Thus both possibilities can be realised.

The 3 possibilities of occupying 2 $O^-$-planes are related by $SL(2,\Z)$, as are the 3 possibilities of occupying 4 $O^-$ planes. Only one of each set of possibilities solves the boundary conditions (\ref{O4n}) (as these are not $SL(2, \Z)$ invariant), and actually, the two possibilities are each others $\Z_2$ complement, as before.

We therefore have two orientifolds representing $SO(4k)$-theory on a 3--torus with spin- nor vector structure. Each orientifold background is of the form $T^3/\Z_2$, with 6 $O^-$ planes, and 2 $O^+$-planes. One orientifold has 2 $O^-$ planes occupied, and the other has the remaining 4 $O^-$-planes occupied.  

$SO(4)$ with spin- nor vector structure has only 2 branes on the dual orientifold, so only one out of the two possibilities mentioned can be realised. This corresponds to 4 components on the moduli space, all consisting of a single point. This can be seen as follows. $Spin(4) = SU(2) \times SU(2)$, and therefore we may write $Spin(4)$ holonomies as $SU(2)$-pairs. The holonomies obeying the boundary conditions are (up to conjugation)
\be
\Omega_1 = (\im \sigma_3, \im \sigma_3)  \qquad \Omega_2 = (\pm \im \sigma_1, \pm \im \sigma_3)  \qquad \Omega_3 = (\im \sigma_1, \im \sigma_1)
\ee
There are four inequivalent possible choices for the signs in $\Omega_2$ (one cannot change a sign in $\Omega_2$ by conjugation without changing some sign in the other holonomies).

For $SO(4k)$ $(k>1)$ and larger groups each of the two orientifold descriptions represents two components. Each of the two orientifold descriptions again represents 2 components. For $SO(4k)$, the two components contribute $k$, resp. $k-1$ to the Witten index. Taking into account the correct multiplicities, gives the correct answer $4k-2$ for the Witten index \cite{Borel}.

\section{Conclusions}

We have shown how to construct orientifold configurations describing compactifications of gauge theories with classical groups on a 2-- or 3--torus. Conversely these results show how various orientifolds should be interpreted as gauge theories. The methods can be extended to higher dimensional tori.

Notice that for two dimensional orientifolds $T^2/\Z_2$ any configuration of $O^+$ and $O^-$-planes is possible. For 3--dimensional orientifolds $T^3/\Z_2$ we only find even numbers of $O^+$ (and $O^-$) planes. There is a simple argument why odd numbers of $O^+$ planes are not allowed \cite{deBoer}.

Some of these configurations are of immediate interest for consistent string-theories. We already mentioned that the $U(n)$-theory on a 2--torus, with holonomy with outer automorphism has precisely the same background as a certain IIB orientifold, though consistency requires the absence of D-branes, and hence the absence of gauge symmetry (so $n=0$ in a sense). For the type I string theory on a 3--torus and its duals, which are argued to have $Spin(32)/\Z_2$ as its gauge group, it seems that there are four configurations of interest, two describing periodic boundary conditions, and two describing absence of vector structure. In a subsequent publication \cite{deBoer} it will be argued however that two of the four configurations actually suffer a relatively subtle inconsistency, invalidating the naive application of group theory methods in string theory. 

{\bf Acknowledgements}: We would like to thank Pierre van Baal and Jan de Boer for useful discussions and reading a draft version of this article.

\end{document}